\documentclass[useAMS,usenatbib,usegraphicx]{mn2e}

\title{Do galaxies form a spectroscopic sequence?}

\author[Y.~Ascasibar and J.~S\'{a}nchez~Almeida]
{
Y.~Ascasibar$^1$ and J.~S\'{a}nchez~Almeida$^{2}$\\
$^1$Departamento de F\'{i}sica Te\'{o}rica, Universidad Aut\'{o}noma de Madrid, Madrid 28049, Spain\\
$^2$Instituto de Astrof\'{i}sica de Canarias, E-38205 La Laguna, Tenerife, Spain
}

\date{{\bf Draft version 2.1} (\today)}

\newcommand{\be}{\begin{equation}}
\newcommand{\ee}{\end{equation}}
\newcommand{\bea}{\begin{eqnarray}}
\newcommand{\eea}{\end{eqnarray}}

\usepackage{color}

\begin{document}

\maketitle

\begin{abstract}
We identify a spectroscopic sequence of galaxies, analogous to the Hubble sequence of morphological types, based on the Automatic Spectroscopic K-means (ASK) classification.
Considering galaxy spectra as multidimensional vectors, the majority of the spectral classes are distributed along a well defined curve going from the earliest to the latest types, suggesting that the optical spectra of normal galaxies can be described in terms of a single affine parameter.
Optically-bright active galaxies, however, appear as an independent, roughly orthogonal branch that intersects the main sequence exactly at the transition between early and late types.
\end{abstract}


\section{Introduction}

It is well known that galaxies can be classified into a small number of morphological types, arranged into a well-defined sequence.
The scheme proposed by \citet{Hubble26}, still widely in use today, is based on the optical appearance of galaxy images, and it divides the galaxy population into ellipticals, lenticulars, and spirals, with the irregular class encompassing all the objects that do not fit into any of the other categories.
It focuses on the symmetry of the galaxy, the concentration of the light towards the centre, and the presence of other features such as disks, bars, and spiral arms.
The Hubble sequence, also known as the Hubble tuning fork, smoothly connects the different morphological classes.
Ellipticals, regular spirals, and barred spirals occupy three different arms of the sequence, intersecting at the lenticular class.
Irregular galaxies are more difficult to accommodate, but they are usually placed at the end of the spiral branches.

The Hubble sequence correlates with the colours of the galaxies, with ellipticals tending to be red and spirals tending to be blue \citep[][]{hum31,hub36,mor57}. 
The relationship, however, presents a large scatter \citep[][]{Connolly+95,sod97,fer06}; about half of the red galaxies are actually disks \citep[][]{mas10,Sanchez-Almeida+11}, and blue ellipticals are not so rare as one may naively think \citep[e.g.,][]{sch09b,hue10a}.
Many works have tried to relate the spectral energy distribution (SED) of a galaxy to its position along the Hubble sequence.
The results are varied, and they are probably affected by the scatter of the relationship between morphological type and spectroscopic class.
\citet{mor57} assigned the blue part of the visible spectrum to stellar classes from A to K, finding a clear relationship in the vein mentioned above.
\citet{aar78} shows how the visible and IR colours of galaxies along the Hubble sequence can be understood as a one-parameter family, in terms of the superposition of spectra of A0V dwarf stars and M0III giants.
\citet{ber95} points out that a simple model consisting of two stellar spectral types can reproduce the observed broad-band colours, but only if the spectral types are
allowed to vary, which implies that the family is not one-dimensional.
Similar conclusions are also reached by \citet{zar95} using stellar spectrum fitting.

One of the most popular spectral classification methods is Principal Component Analysis (PCA).
It is fast and robust, and it has a sound mathematical foundation \citep[see e.g.][]{Everitt95}.
For a given dataset, PCA finds the smallest possible set of orthogonal eigenvectors that reproduce the data within a certain accuracy.
In the case of galaxy spectra, \citet{Connolly+95} claim that the first two eigencoefficients suffice to represent most galaxies, and that the resulting spectral types can be described in terms of a one-parameter family.
Based on the PCA decomposition of a much larger galaxy sample, containing more than $10^5$ spectra, \citet{Yip+04} highlight the importance of the third eigenvector and summarize the galaxy spectra in terms of two independent angular variables.
In other words, the galaxies are contained within a three-dimensional volume, given by the linear combination of the first three eigenvectors.

However, it is not obvious whether the galaxy distribution is indeed three-dimensional, or it is confined to a non-linear manifold of lower dimensionality, immersed in the three-dimensional space.
Here we study the multidimensional distribution of galaxy spectra and explore the possibility that different galaxy types may be arranged into a spectroscopic sequence, analogous to the morphological Hubble tuning fork.
The first problem, of course, is how to detect such a sequence, if it existed, in a space with as many dimensions as data points in the spectrum.
Then, if galaxies did indeed form a well-defined spectroscopic sequence, it would be extremely interesting to quantify its multidimensional structure.
Would all galaxies be arranged along a single curve? along several branches? along a hyperplane (or a higher-dimensionality subspace)?

In principle, the dimensionality of the sequence is related to the number of parameters that are necessary in order to fully describe a galaxy spectrum.
If all their observable properties depended on only one single degree of freedom, galaxies would describe a one-dimensional curve in spectral space, no matter how complicated.
For two parameters, the galaxy population would define a `fundamental hypersurface' (not necessarily a plane), and so on.
It must be noted, though, that these subspaces may or may not be fully occupied.
Galaxies could be arranged in several disconnected clumps, more or less randomly distributed in spectral space, or be confined to a certain region defined by some set of inequalities.

\begin{figure*}
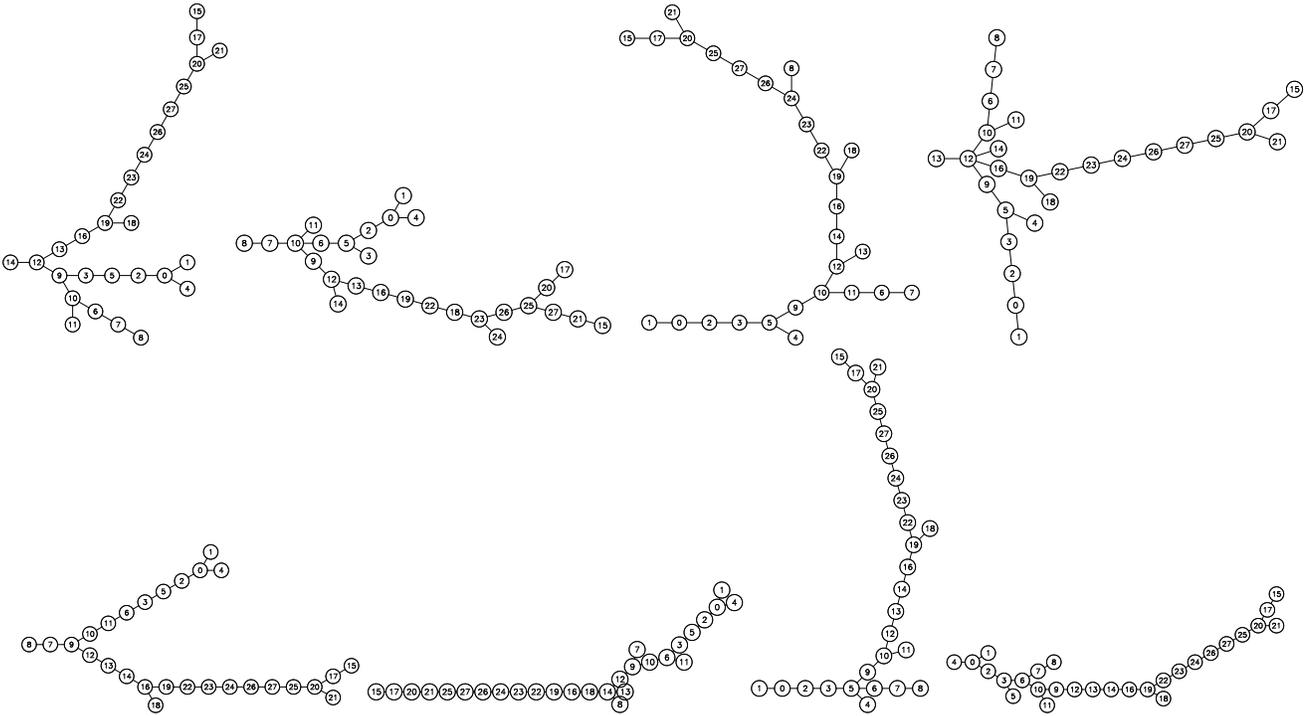

\includegraphics[width=3cm]{figs/ask_mst_dist.eps}
\includegraphics[width=5cm]{figs/ask_mst_red.eps}
\includegraphics[width=4cm]{figs/ask_mst_blue.eps}
\includegraphics[width=5cm]{figs/ask_mst.eps}

\includegraphics[width=4.5cm]{figs/manhattan_mst_dist.eps}
\includegraphics[width=5cm]{figs/manhattan_mst_red.eps}
\includegraphics[width=2.5cm]{figs/manhattan_mst_blue.eps}
\includegraphics[width=4.5cm]{figs/manhattan_mst.eps}
\caption
{
Minimal spanning tree of the ASK classes, based on the Euclidean (top) and Manhattan (bottom) distance between template spectra, considering, from left to right, the full wavelength range, only the red part ($\lambda>6000$~\AA), the blue part ($\lambda<6000$~\AA), and the same bands used to define the ASK classes.
}
\label{figMST}
\end{figure*}

In practice, finding, let alone characterizing, a non-linear two-dimensional hypersurface is by no means a trivial task, and even more so for structures in many dimensions \citep[see e.g.][]{AscasibarBinney05,Ascasibar08,Ascasibar10}.
However, the existence of several relations, like the \citet{TullyFisher77} relation for spiral galaxies, or the \citet{FaberJackson76} relation and the fundamental plane \citep{DjorgovskiDavis87} for elliptical galaxies, provide encouraging evidence that galaxies can be described in terms of very few independent parameters \citep{Disney+08, Tollerud+_10}.
As mentioned above, previous studies based on principal component analysis have concluded that only the first two or three linear eigencoefficients are necessary in order to reproduce the main features of the galaxy distribution in spectral space \citep[e.g.][]{GalazLapparent98,Connolly+95,Castander+01,Yip+04,Ferreras+06}.

The present work follows a different approach, based on the publicly-available\footnote{{\tt ftp://ask:galaxy@ftp.iac.es/}} Automatic Spectroscopic K-means-based (ASK) classification of all the galaxy spectra in the seventh data release of the Sloan Digital Sky Survey \citep[SDSS/DR7,][]{Stoughton+02,Abazajian+09}.
A thorough description of this classification scheme is provided in \citet{Sanchez-Almeida+10}, where the reader is referred to for further details, but the main aspects are summarized in Section~\ref{askclasses} for the sake of comprehensiveness.
We investigate whether galaxies form a continuous, single-parameter sequence by studying the minimal spanning tree (MST) of the template vectors defining the ASK classes.
The details of the computation of the minimal spanning tree are given in Section~\ref{secMST}, and Section~\ref{secSequence} is devoted to the identification of a possible spectroscopic sequence.
A quantitative characterization and its physical interpretation are discussed in Section~\ref{secDiscussion}, and our main conclusions are then succinctly summarized in Section~\ref{secConclusions}.

\section{The ASK classification}
\label{askclasses}

The ASK classes are the result of classifying all the galaxies with spectra in the SDSS/DR7.
Those with redshift smaller than 0.25 are transformed to a common rest-frame wavelength scale, and then re-normalized to the integrated flux in the SDSS $g$-filter.
These two are the only manipulations the spectra undergo before classification.
The classification is driven only by the shape of the spectra, and these two corrections remove the obvious undesired dependencies on the redshift and apparent magnitude of the galaxy.
\citet{Sanchez-Almeida+10} deliberately avoid correcting for other known effects requiring modelling and assumptions (e.g., dust extinction, seeing, or aperture effects), in the spirit of the rules for a good classification put forward by \citet{Sandage05}, where it is pointed out that physics must not drive a classification.
Otherwise, the arguments become circular when the classification is used to infer the underlying physics.

The classification algorithm used, k-means, is a well-known, robust workhorse, commonly employed in data mining, machine learning, and artificial intelligence \citep[see e.g.][]{Everitt95,Bishop06}.
Its computational efficiency was an important asset in order to carry out the simultaneous classification of the full data set ($\sim 12$~GB).
The algorithm works by iteratively assigning each galaxy to the nearest class in spectral space and re-evaluating the class template spectrum as the average over all the class members.
In the end, 99\% of the galaxies can be assigned to only 17 major classes, with 11 additional minor classes describing the remaining one percent.
The actual number of classes has some uncertainty, although it is automatically provided by the algorithm, which always renders between 15 and 19 major classes.
The template spectra vary smoothly and continuously, and they are labelled from 0 to 27 according to their $(u-g)$ colour, from reddest to bluest.
It is unclear whether the ASK classes represent genuine clusters in the 1637-dimensional classification space, or they partake a continuous distribution 
\citep[see the discussion in][as well as Section~\ref{secDiscussion}]{Sanchez-Almeida+10}.
The class templates cover all the possible spectral shapes, and we use them in our search for a sequence.
Since we are not interested in the occupation distribution along such a sequence, all templates are treated equally, even though each class contains a different number of SDSS/DR7 galaxies.

\section{Minimal spanning tree}
\label{secMST}

The MST of a graph \citep[e.g.][]{Kruskal56} is the set of edges that connect all the vertices in the graph at a minimum cost, defined as the sum of the individual costs of all the edges included in the tree.
In our case, these individual costs are given by the differences between the template spectra, and the MST can be thought of as the shortest possible `road network' connecting all spectral classes.

Although real life is a little bit more complicated (see below), one may expect that, if galaxies, and thus classes, were roughly arranged into a single curved line, the MST would be ideally suited to identify such a multidimensional sequence.
For example, if A, B, C, D, and E are different types of galaxy, forming the sequence A-B-C-D-E, this would be their minimal spanning tree, and it can be proven that any other combination of edges would result in a longer total distance.
In this example, the extreme classes A and E would have just one connection, whereas the intermediate types B, C, and D would have two.
Moreover, the distance between next-to-consecutive classes (e.g. A and C) must be larger than both AB and BC.
If A, B, and C are aligned along a straight line, the equality ${\rm AC = AB + BC}$ will hold, whereas for a curved line ${\rm AC < AB + BC}$.

If the galaxy distribution had more than one dimension, some classes would become `tree nodes' featuring three or more connections.
If the subspace defined by the galaxies is fully occupied, there will be a large number of nodes, and it would be difficult to obtain much information about its structure from the MST alone.
On the other hand, a small number of nodes would imply that galaxies are arranged into a few discrete `branches' with different orientations.

The minimal spanning tree of ASK classes is shown in Figure~\ref{figMST}, using eight slightly different definitions of the distance in spectral space.
In the top panels, we assume the Euclidean metric
\be
d_{AB}^2 = \sum_{i=1}^{N_\lambda} \left[\, C_B(\lambda_i) - C_A(\lambda_i) \,\right]^2
\label{ecDistance}
\ee
whereas the Manhattan distance
\be
d_{AB} = \sum_{i=1}^{N_\lambda} \left|\, C_B(\lambda_i) - C_A(\lambda_i) \,\right|
\ee
has been used in the bottom panels.
In both cases, $d_{AB}$ denotes the distance between classes $A$ and $B$, and $C_X(\lambda)$ is the spectral template of class $X$, evaluated at a wavelength $\lambda$.
We investigate different criteria to select the set of $N_\lambda$ discrete wavelengths involved in the computation in order to test the stability of the results: the whole spectral energy distribution between 3800 and 9250~\AA\ ($N_\lambda=3850$), only the bluest part ($\lambda<6000$~\AA, $N_\lambda=1977$), the red part ($\lambda>6000$~\AA, $N_\lambda=1873$), and the same 17 bands that were used to define the ASK classification \citep[$N_\lambda=1637$, see Table~1 in][for the precise definition of the bandpasses]{Sanchez-Almeida+10}.
Note that, in each case, the dimensionality of the data space is given by the value of $N_\lambda$.

\begin{figure}
\centering
\includegraphics[width=8cm]{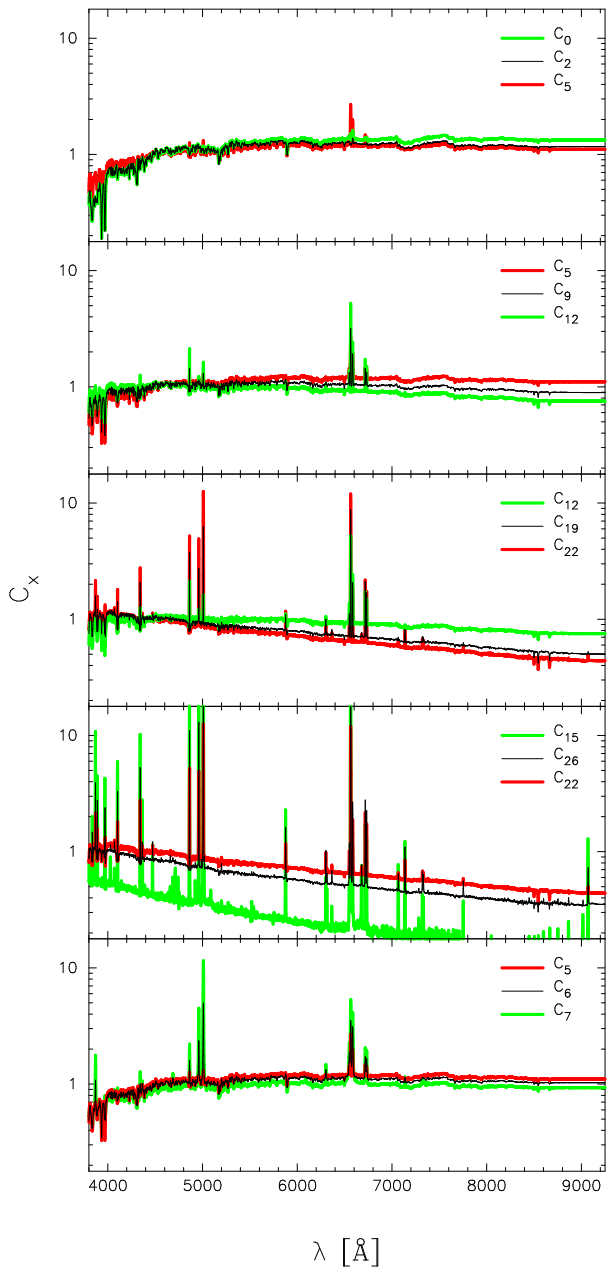}
\caption
{
Template spectra of several ASK classes representative of the early-type (top panel), late-type (middle panels), and active (bottom panel) branches.
$C_X$ denotes the spectral template of class $X$.
}
\label{figSEDs}
\end{figure}

The fourth option -- using the spectral range used to define the ASK classification -- is arguably the most natural.
Using the spectral range to the red end, where the mean spectra of each class have been extrapolated, or the blue end, which is often dominated by the presence of strong emission lines, seem to be poor choices that probably introduce some noise in the resulting MST.
Also, the Euclidean metric, where differences add in quadrature, seems to be more appropriate for comparing two spectra than the Manhattan distance.
We simply use all these different definitions in order to test the stability of our results.
Although the solution is far from being unique (the exact ordering of the classes depends on the adopted definition), the overall picture is fairly robust:
according to the MST, the ASK classes representing the galaxy population in the SDSS seem to be distributed along three main spectroscopic branches, or, alternatively, along a main spectroscopic sequence with one ramification.

The longest branch, both in terms of the number of classes and the extent measured by the Euclidean distance (i.e. differences in the template spectra) is composed of ASK types 15, 17, 20, 21, 25, 27, 26, 24, 23, 22, 18, and 19.
The position of these classes on the principal components plane, as well as on the colour-colour and BPT \citep{Baldwin+81} diagrams (see Figures~\ref{figDiagrams} and ~\ref{figPCA}) suggests that this branch corresponds to the sequence of dwarf irregular galaxies.
It merges smoothly with the location in spectral space occupied by normal spirals, represented by ASK classes 16, 14, 13, 12, and 9.
Early-type galaxies (classes 0, 2, 3, and 5) are also grouped together in another branch for all measures of the distance, and the same can be said of the active galaxy types 8, 7, 6, 11, and 10.
All three branches (early-type, late-type, and active) seem to converge around classes 9, 10, or 12, depending on the definition used.
Classes 1 and 4 seem to be outsiders in the main red sequence, and we are currently investigating the possibility that they are associated to heavily dust-reddened spirals \citep{Sanchez-Almeida+11}.
A sample of spectra that are representative of each branch are plotted in Figure~\ref{figSEDs}.

\section{A spectroscopic sequence?}
\label{secSequence}

We argue that the three independent branches we have identified trace an underlying spectroscopic sequence, analogous to the Hubble tuning fork of galaxy morphologies, and the subtle differences between the MST obtained for different definitions of the distance are due to the presence of random deviations of the individual galaxy spectra with respect to the average behaviour of the sequence.

In other words, our branches are not ideally thin hyperlines in the data space, but `hypertubes' with a certain, variable thickness, where the contributions of intrinsic physical dispersion of the galaxy properties as well as extrinsic observational errors add in quadrature.
Given the large number of objects involved in most ASK classes, measurement errors have a negligible effect on the mean spectrum, but they make a significant contribution to the dispersion of the individual galaxies around the mean (although the error in the mean spectrum decreases as the square root of the number of observations, the actual dispersion of the distribution is independent of the number of galaxies).

\begin{figure}
\centering
\includegraphics[width=8cm]{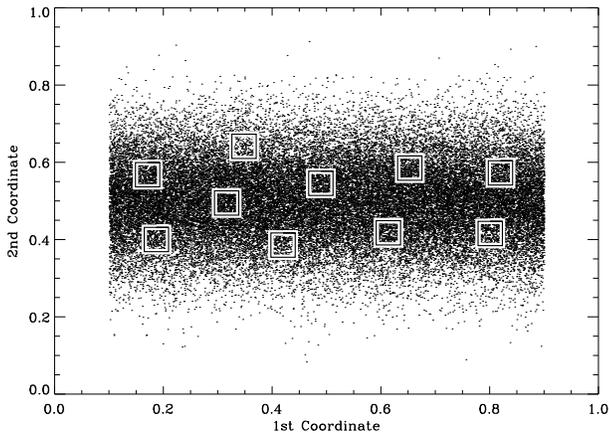}
\caption
{
Results of the k-means algorithm for a random distribution of points in two dimensions (see text).
Class centres are shown by the open boxes.
}
\label{figKmeans}
\end{figure}

Due to the finite thickness of the branches, the ASK classes derived from the k-means algorithm will not be aligned along the centres of these hypertubes, but they will alternate along their boundaries.
This can be easily illustrated by a simple experiment, where we set up a random distribution of data points that corresponds to a hypertube in two dimensions.
The first coordinate varies uniformly from 0.1 to 0.9, and the second follows a Gaussian distribution centered at 0.5 with a standard deviation of 0.1.
The data points and the centres of the final classes returned by the k-means algorithm are plotted in Figure~\ref{figKmeans} as dots and open boxes, respectively.
This configuration, where classes (i.e. template spectra) alternate between the boundaries of the distribution rather than tracing the centre, will occur whenever the dispersion around the mean is comparable to the typical inter-class distance.
During the first iteration, the classes are initialized at the locations of randomly-picked data points.
Then, each class collects the points in its Voronoi cell, and its position is updated to the new centre of mass.
The process is repeated until the classes arrange themselves into the pattern shown in Figure~\ref{figKmeans}, which roughly corresponds to the most efficient packing in two dimensions.

\begin{figure*}
\includegraphics[width=\textwidth]{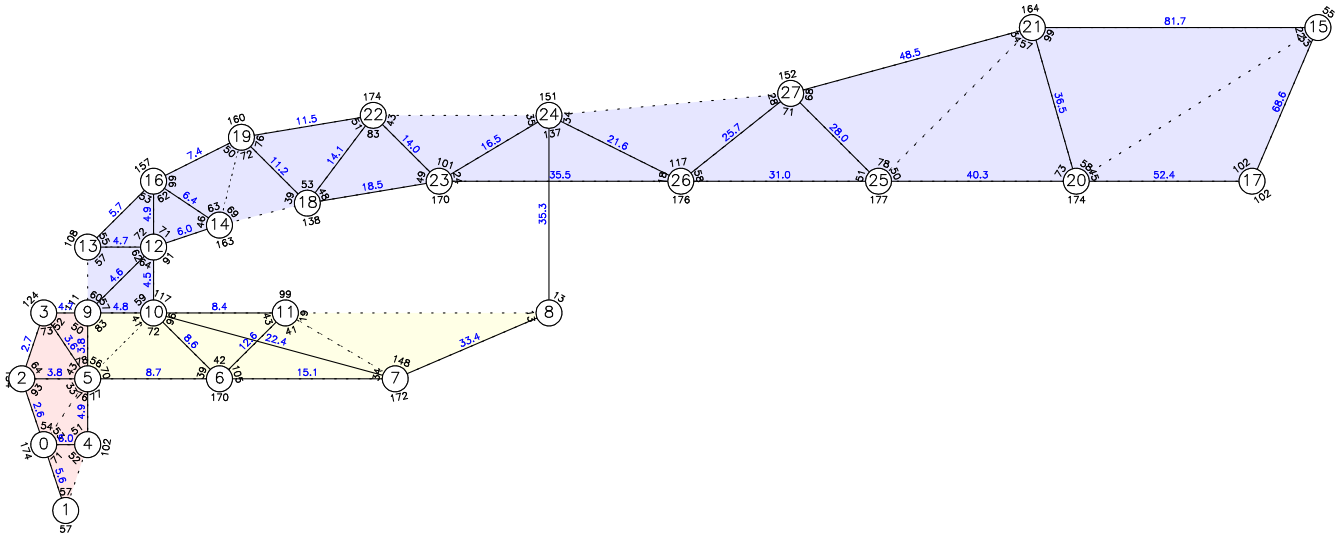}
\caption
{
Distances (blue) and angles (black) between all the edges that are present in any of the four MST depicted on the top panel in Figure~\ref{figMST}.
Solid lines show all the connections appearing at least once, for any definition of the distance, but numeric values have been computed using the same bands as in the definition of the ASK classes.
The early, late, and active branches have been highlighted in different shades, and dotted lines have been added to guide the eye when reading the angles.
}
\label{figSequence}
\end{figure*}

In the general case, the MST will be able to pinpoint a sequence with a finite thickness, but it will zigzag through the distribution rather than crossing it along a more or less straight line.
More precisely, the angle between the directions of two consecutive edges (say, AB and BC) may not necessarily be small, and the distance AC may be comparable to both AB and BC.
On the contrary, the segments AC, CE, EG, etc. -- which do \emph{not} belong to the MST -- trace the boundary of the distribution and, unless there is a sharp turn, the angles between them will be typically smaller than the angles between consecutive edges in the MST.

We therefore investigated all the distances and angles between consecutive vertices in any of the four variants of the MST defined with the Euclidean metric distance shown in the previous section.
A schematic representation is plotted in Figure~\ref{figSequence} (see also Appendix~\ref{secFigure}).
Being a projection of a non-linear multidimensional structure, it is necessarily not to scale, but we have tried to reproduce actual distances and angles as faithfully as possible without sacrifying clarity.
Precise values of the distances and angles, based on the ASK bands (last MST in Figure~\ref{figMST}) are indicated by the small numbers in blue and black colour, respectively.
Distances are given by equation~(\ref{ecDistance}), while the angle between two vectors is computed from the usual Euclidean scalar product
\be
\cos\alpha = \frac
{ \sum_{i=1}^{N_\lambda} v_{AB}(\lambda_i)\, v_{AC}(\lambda_i) }
{\sqrt{ \sum_{i=1}^{N_\lambda} v_{AB}(\lambda_i)^2\ \sum_{i=1}^{N_\lambda} v_{AC}(\lambda_i)^2 }}
\label{eqAngles}
\ee
where
\be
v_{AB}(\lambda_i) = C_B(\lambda_i) - C_A(\lambda_i)
\ee
is the vector connecting the spectral templates of classes $A$ and $B$, and a similar definition holds for $v_{AC}$.
Although the precise values of the distances depend, of course, on the adopted wavebands, neither the relations between them nor the angle between two edges are very sensitive to that choice.

In order to illustrate the figure, let us take a simple node, for example number 21.
There is one edge towards node 20 because these two classes are connected in the first, third, and fourth panels in Figure~\ref{figMST}, and there are two additional edges towards 15 and 27 because of the connections in the second panel.
The distances from class 21 to classes 15, 20, and 27 are marked in blue as 81.7, 36.5, and 48.5, respectively.
The angles between edges $21-27$ and $21-25$ (34 degree), $21-25$ and $21-20$ (57 degree), $21-20$ and $21-15$ (99 degree), and $21-15$ and $21-27$ (164 degree) are indicated by the black numbers.

Our results are consistent with the pattern described above, where ASK classes would be arranged along the boundaries of three independent branches with a thickness of about two classes.
These branches are not straight lines, nor do they lie in the same hyperplane, but they represent clearly defined sequences in spectral space.
The angles between the edges that trace the boundaries are close to 180 degrees, indicating that the sequences describe a relatively smooth curve, while the interior angles are close to 60 degrees, implying the classes are arranged in roughly equilateral triangles within the sequence.

\section{Discussion}
\label{secDiscussion}

Our main result is the identification of a spectroscopic sequence with three separate branches, which represent -- at least in a qualitative sense -- early-type, late-type, and active galaxies.
However, several questions remain open:
How many parameters are necessary in order to fully specify the spectral properties of a galaxy?
How could one compute their value, and what is their physical meaning?
What do they tell us about galaxy formation and evolution?

Concerning the first question, the configuration of the ASK classes in spectral space suggests that the optical spectra in the SDSS/DR7 have only two degrees of freedom.
An interpretation in terms of a single parameter would be that the putative one-dimensional curve in spectral space marches through the early-type galaxies, climbs up and down the active branch, and then moves on towards the late types.
Alternatively, and arguably more likely, it would also be possible that there is a main sequence going from early- to late-type galaxies, but some of them \citep[especially those in the green valley; see e.g.][]{Salim+07} may be temporarily found in an active state that takes them out of the main sequence.
In this scenario, the optical spectrum of a galaxy can be accurately described in terms of one discrete parameter (whether the galaxy belongs to the `normal' or the `active' branches) and one real number characterizing its position along the corresponding sequence.

This interpretation is reinforced when one tries to find a bidimensional projection that captures the main features of the spectral classification.
After some experimentation, we have selected the hyperplane defined by ASK classes 0, 5, and 8, corresponding to a typical early-type galaxy, an object in the green valley, and an extremely active galaxy, respectively.
Looking at the distances and angles between the classes, one can easily verify that classes 5, 6, 7, and 8 are roughly aligned along a more or less straight line, at an angle of 102 degrees (almost perpendicular) with respect to the segment connecting classes 5 and 0.
We have thus selected the template of the green-valley class 5 as the origin of coordinates.
The axis towards class 8 provides a measure of galaxy activity, and we have set the normalization so that this class represents the unit value.
For a galaxy with spectral energy distribution $S(\lambda)$, the activity $A$ is defined as the scalar product
\be
A = \sum_{i=1}^{N_\lambda} \left[ S(\lambda_i) - C_5(\lambda_i) \right] \Delta_A(\lambda_i)
\ee
with the vector
\be
\Delta_A(\lambda_i) =
\frac{ C_8(\lambda_i) - C_5(\lambda_i) }
     { \sum_{j=1}^{N_\lambda} \left[ C_8(\lambda_j) - C_5(\lambda_j) \right]^2 }
\ee
On the other hand, the axis defined by class 0 can be used to quantify the spectral type of the galaxy in a similar way:
\be
T = \sum_{i=1}^{N_\lambda} \left[ S(\lambda_i) - C_5(\lambda_i) \right] \Delta_T(\lambda_i)
\ee
where
\bea
\nonumber
\Delta_T(\lambda_i) &\propto& \left[ C_0(\lambda_i) - C_5(\lambda_i) \right] \\
& & -
\frac{ \sum_{j=1}^{N_\lambda} \left[ C_0(\lambda_j) - C_5(\lambda_j) \right] \Delta_A(\lambda_j) }
     { \sum_{j=1}^{N_\lambda} \Delta_A^2(\lambda_j) }
\Delta_A(\lambda_i)
\label{eqDeltaT}
\eea
is orthogonal to $\Delta_A$, and it is normalized so that the type of class 0 is equal to $-1$, i.e., we chose the scaling constant in equation~(\ref{eqDeltaT}) so that
\be
\sum_{i=1}^{N_\lambda} \left[ C_0(\lambda_i) - C_5(\lambda_i) \right] \Delta_T(\lambda_i) = -1
\ee

\begin{figure}
\centering
\includegraphics[width=8cm]{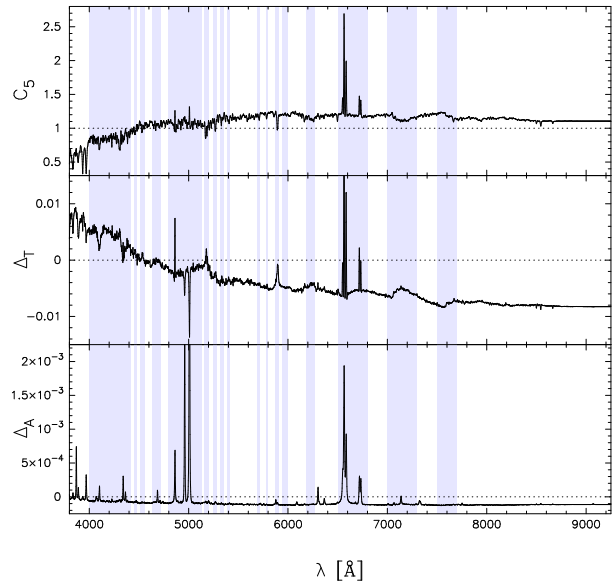}
\caption
{
Basis for the galaxy type-activity decomposition.
The coordinate origin, set by the template spectrum of ASK class 5, is shown on the top panel.
Middle and bottom panels show the basis vectors $\Delta_T$ and $\Delta_A$ determining type and activity, respectively.
The blue bands indicate the ASK bandpasses, also used to compute the values of type and activity plotted in Figure~\ref{figPlane}.
}
\label{figBasis}
\end{figure}

\begin{table}
\begin{center}
\begin{tabular}{ccccc}
\hline
$\lambda$ [\AA]& $C_5$  & $\Delta_T$ & $\Delta_A$ & mask \\
\hline
    3800.00 & 4.689e-01 & 3.749e-03  &  3.218e-05 & 0 \\
    3800.88 & 4.986e-01 & 4.272e-03  &  1.624e-05 & 0 \\
    3801.76 & 5.361e-01 & 4.962e-03  & -5.754e-06 & 0 \\
    3802.64 & 5.724e-01 & 5.554e-03  & -1.339e-05 & 0 \\
    3803.51 & 6.024e-01 & 6.045e-03  & -2.552e-05 & 0 \\
    3804.39 & 6.226e-01 & 6.356e-03  & -3.207e-05 & 0 \\
    3805.27 & 6.319e-01 & 6.480e-03  & -3.243e-05 & 0 \\
    3806.15 & 6.324e-01 & 6.466e-03  & -3.256e-05 & 0 \\
    3807.03 & 6.310e-01 & 6.484e-03  & -2.825e-05 & 0 \\
\ldots & \ldots & \ldots & \ldots & \ldots \\
\hline
\end{tabular}
\end{center}
\caption
{
Template spectrum of ASK class 5 and basis vectors $\Delta_T$ and $\Delta_A$ as a function of wavelength.
Last column is a binary flag indicating whether that particular wavelength is included in the bands used to define the ASK classes.
The full table is available in the electronic version.
}
\label{tabBasis}
\end{table}

\begin{figure}
\centering
\includegraphics[width=8cm]{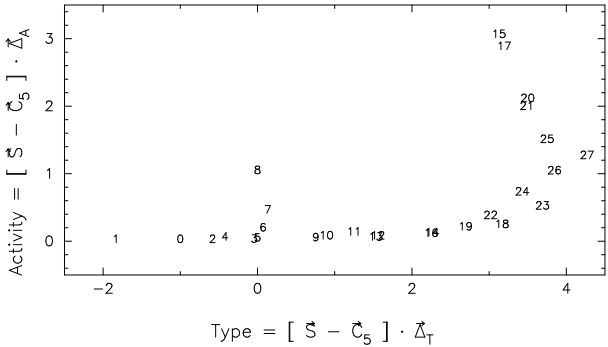}
\caption
{
Location of ASK classes in the galaxy type-activity plane.
}
\label{figPlane}
\end{figure}

The spectral template of class 5, together with the basis vectors $\Delta_A$ and $\Delta_T$, are represented in Figure~\ref{figBasis}, and the location of ASK classes in the galaxy type-activity plane is plotted in Figure~\ref{figPlane}.
The shape of the spectroscopic sequence is evident in this representation, which has the advantage that its two parameters are easy to evaluate for any galaxy (numeric values for $C_5$, $\Delta_A$, and $\Delta_T$ are provided in Table~\ref{tabBasis}) and have a pretty clear physical meaning: the spectral type quantifies the evolutionary state of the galaxy, while the activity component reflects the presence of warm ionized gas.
An interesting feature that was not obvious from the study of the angles between edges in the MST is that the end of the late-type branch turns into a direction that is roughly parallel (more precisely, forms an angle of 20 degree) in spectral space to the axis defined by the active galaxies.
From Figure~\ref{figBasis} we immediately see that this is due to the presence of strong emission lines that make a significant contribution to the overall luminosity, but both branches can be easily identified with the well-known AGN/starburst dichotomy in the BPT diagram.

\begin{figure}
\centering
\includegraphics[width=8cm]{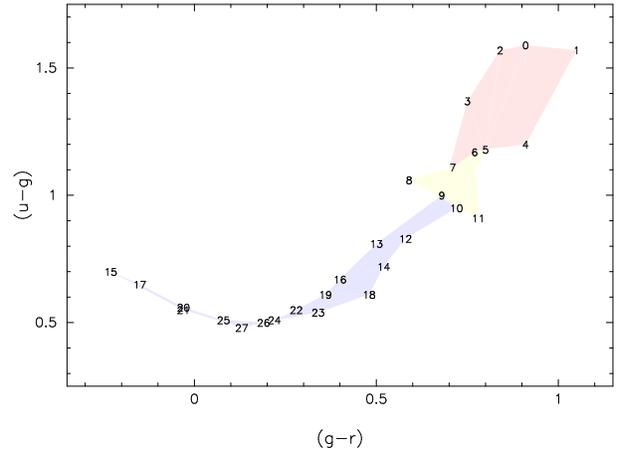}\\[5mm]
\includegraphics[width=8cm]{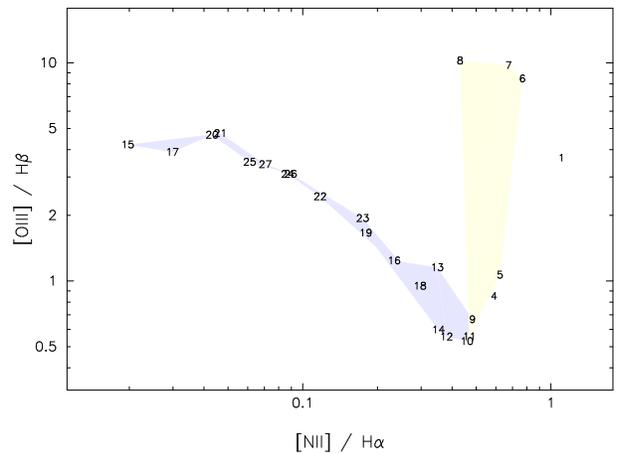}
\caption
{
The spectroscopic sequence in the colour-colour (top) and BPT (bottom) diagrams.
}
\label{figDiagrams}
\end{figure}

The projection of our spectroscopic sequence onto the BPT diagram, as well as the colour-colour plane defined by $(u-g)$ and $(g-r)$, is shown in Figure~\ref{figDiagrams}.
The early and late branches correspond to the red sequence and the blue cloud in the colour-colour plot, respectively, and all the active galaxies are located approximately at the same place, in the region of the green valley.
On the other hand, AGN can be neatly separated from the late-type galaxy population in the BPT diagram, but in this case the early-type branch is absent because the intensity of the H$\beta$ line is too faint (in fact, it is observed in absorption for ASK classes 0, 2, and 3).
The projection onto the type-activity plane has the advantage that the full structure of the sequence, with its three branches, can be represented simultaneously.

\begin{figure}
\centering
\includegraphics[width=8cm]{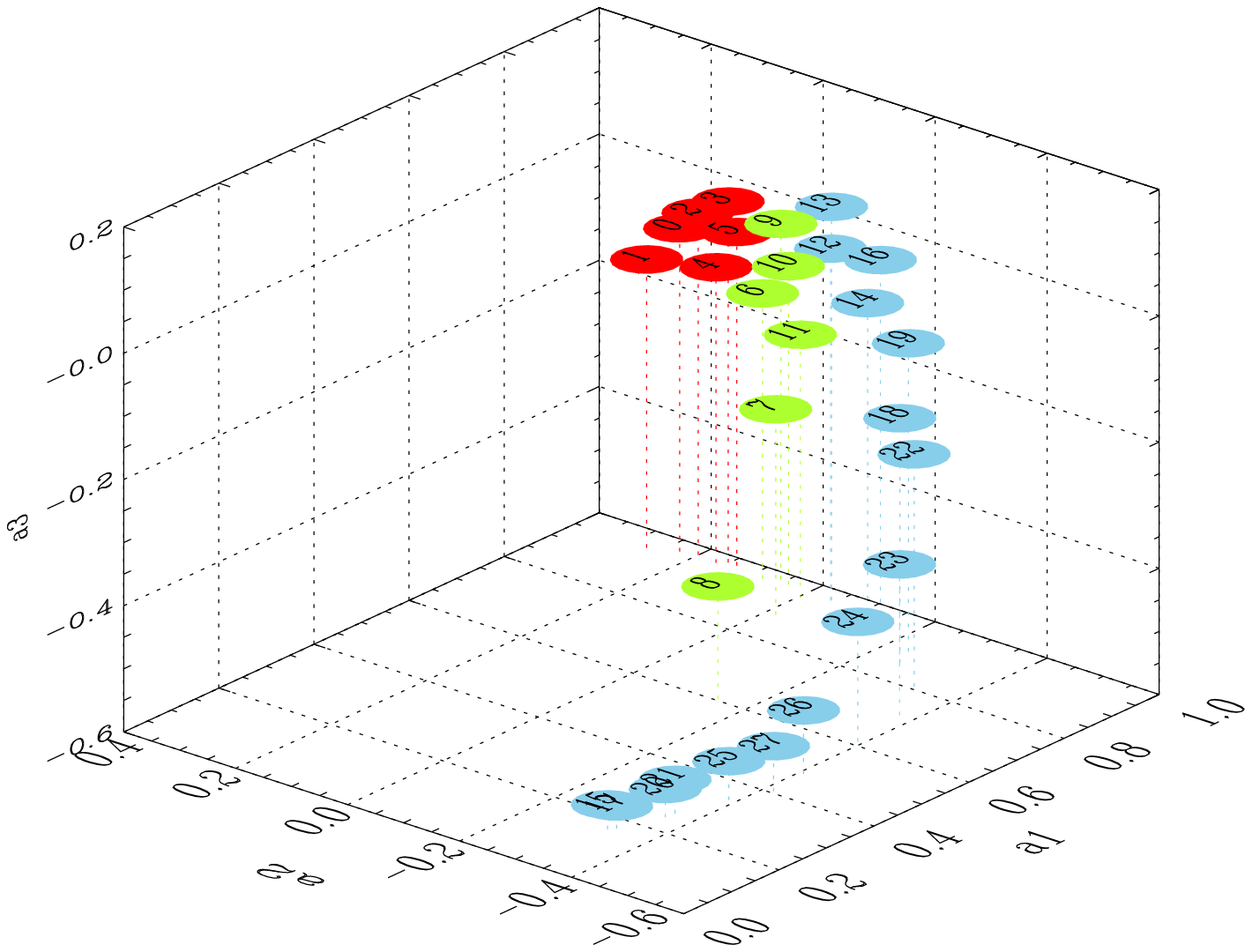}\\
\includegraphics[width=8cm]{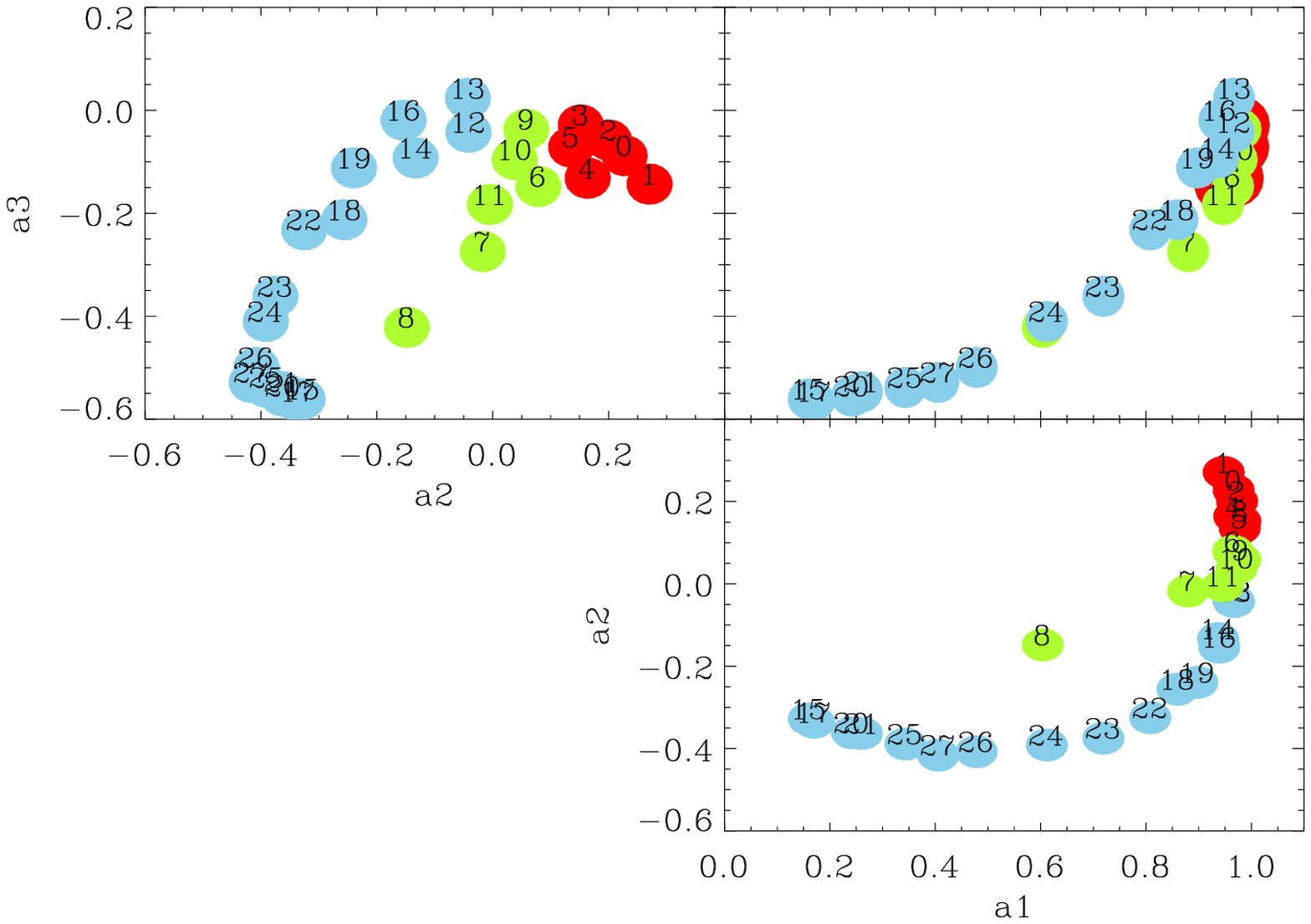}
\caption
{
Top: Distribution of the ASK classes in the space of the first three eigenvalues of the PCA decomposition by \citet{Yip+04}.
Note how the classes follow a one-dimensional curve in this three-dimensional space, with a diverticulum corresponding to the active branch.
Bottom: Projection of the curve in the three orthogonal planes.
}
\label{figPCA}
\end{figure}

Another interesting space for projection is the volume defined by the first three eigenvectors\footnote{Retrieved from {\tt http://www.sdss.org/}} resulting from a principal component analysis \citep{Yip+04}.
As in the case of the colour-colour and BPT diagrams, the one-dimensional nature of the galaxy distribution is also encoded in the configuration of the ASK classes within the three-dimensional volume defined by the first PCA coefficients $(a_1,a_2,a_3)$, as well as in the projections of the spectroscopic sequence onto the three orthogonal planes $(a_1,a_2)$, $(a_1,a_3)$, and $(a_2,a_3)$ shown in Figure~\ref{figPCA}.
Note that the projection of the curve in the first two eigenvalues $(a_1,a_2)$ is qualitatively similar to the location of the classes in the type-activity plane (Figure~\ref{figPlane}).
This fact connects our sequence with the finding by \citet{Connolly+95} that most galaxy spectra form a one-parameter family defined by the ratio between the first two eigenvalues.
The original PCA-based sequence does not include the active branch we have identified, although it can be guessed in Figure~4 of \citet{Yip+04}.

It is convenient to emphasize that the parameters $A$ and $T$ are not meant to describe the galaxy spectra by themselves.
More precisely,
\be
S(\lambda_i) \neq C_5(\lambda_i) + A \Delta_A(\lambda_i) + T \Delta_T(\lambda_i)
\ee
in contrast to the PCA decomposition.
Their purpose is merely to provide a quantitative, continuous, linear mapping between the SED of a given galaxy and our spectroscopic sequence.
Although a detailed discussion is beyond the scope of the present work, the position of a given galaxy in the type-activity plane may be used for assigning it to the appropriate branch, as well as determining its position along the corresponding one-dimensional curve.
The galaxy spectrum could then be recovered by interpolating the templates of the nearest ASK classes, but the procedure is less straightforward than PCA.

Our results are consistent with other approaches to galaxy classification, but they highlight the remarkably few degrees of freedom that are necessary in order to characterize optical spectra.
Out of the several thousand dimensions of the data space, the vast majority of SDSS galaxies are confined to just two one-dimensional curves, contained in a three-dimensional Euclidean volume.

While it is well known that the structure of dark matter haloes can be described in terms of one or two free parameters \citep[see e.g.][and references therein]{AscasibarGottloeber08}, related to the statistical properties of the primordial perturbations of the density field \citep[e.g.][]{Ascasibar+04,Ascasibar+07}, it is somewhat surprising that galaxy formation, with all the complex physical processes involved, does not seem to introduce additional degrees of freedom.
In our opinion, understanding why the optical SED of a galaxy contains so little information is an important piece of the puzzle of galaxy formation and evolution, and it poses a very strong constraint on the number of free parameters that are available to theoretical models.

Finally, let us note that, like the Hubble morphological sequence, the spectroscopic sequence presented here provides a snapshot of the distribution of galaxies in spectral space at the present day, but it does \emph{not} imply an evolutionary track, nor does it contain any temporal information whatsoever.
Galaxies of type 9 are similar to galaxies of type 10, and these, in turn, have spectra that are close to those of galaxies of type 11, but this does not mean that they turn into one another.

Nevertheless, our main sequence can be interpreted in terms of the average age of the stellar population, with younger galaxies corresponding to larger values of the spectral type.
The almost perpendicular location of the active branch with respect to the main sequence suggests that optically-selected AGN are associated with one particular stage of galactic evolution, in agreement with earlier results \citep[e.g.][]{Schawinski+07,Schawinski+10} and consistent with a scenario where quasar activity marks the termination of star formation and the transition from late to early type \citep[see e.g.][and references therein]{Hickox+09}.
If this is true, the type-activity diagram -- and the spectroscopic sequence -- will probably have a similar form at high redshift, although each branch would be located in a different region of the spectral space and populated by galaxies of very different physical properties.

\section{Conclusions}
\label{secConclusions}

In this work, we have investigated the distribution of galaxies in spectral space.
More precisely, we have computed the minimal spanning tree of the class templates in the Automatic Spectroscopic K-means-based (ASK) classification of SDSS/DR7 data.
By studying the distances and angles in spectral space between different galaxy types, it is found that galaxies in the local universe are distributed along a spectroscopic 
sequence with three independent branches.
These branches contain the spectra of early-type, late-type, and active galaxies, and they intersect in the spectral region corresponding to the `green valley'.

This configuration contains two degrees of freedom: one discrete parameter that determines whether a galaxy belongs to the `normal' (either early- or late-type) part of the sequence or to the `active' branch, as well as one continuous affine parameter that describes the position of the galaxy along its one-dimensional branch.
We interpret the normal branches as a main galactic sequence, described in terms of one single affine parameter indicative of the evolutionary state of the object (a combination of mean stellar age, gas abundance, and chemical composition).
At some point in the evolution of a galaxy, star formation stops, and the object moves from the late-type to the early-type branch.
During that phase, some galaxies can also be found in the active branch.
In agreement with previous work, we find that optical AGN activity is exclusively associated with that particular period.

We have verified that our results are robust with respect to different definitions of the distance between galaxy spectra.
A straightforward prescription and a set of basis vectors are provided in order to quickly evaluate the spectral type and activity of a given galaxy.
In the future, we would like to study the relation between spectral type and optical morphology for SDSS galaxies and to apply the same technique to a sample of objects at higher redshifts.

 \section*{Acknowledgments}

The authors would like to thank the referee for a thorough, constructive report that led to significant improvements in the manuscript, as well as A.~I.~D\'{i}az, C.~Jones, W.~Forman, R.~Terlevich, C.~Mu\~noz-Tu\~n\'on and M.~Garc\'{i}a-Vargas for several useful comments and discussions.

This work has been funded by projects AYA\,2007-67965-C03-03, AYA~2007-67965-C03-01, AYA\,2007-67752-C03-01, and CSD\,2006-00070 (Spanish Ministry of Science, Technology and Innovation).

 \bibliographystyle{mn2e}
 \bibliography{references}

\appendix
 \section{Distances and angles between ASK classes}
 \label{secFigure}

\begin{figure}
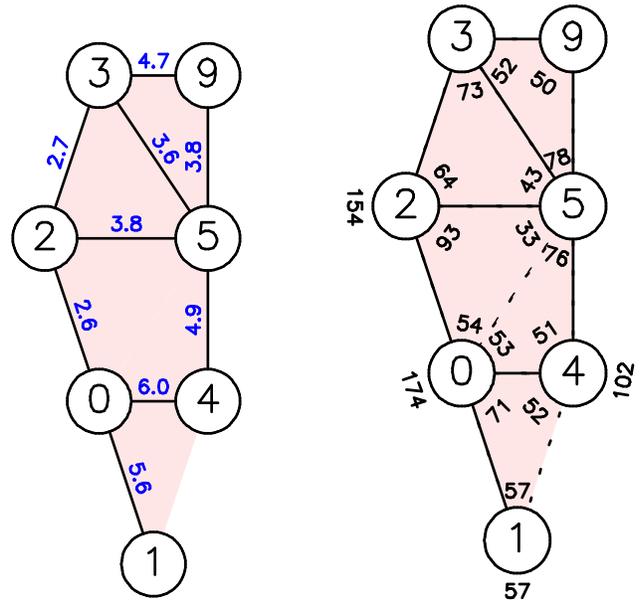

\includegraphics[height=7.5cm]{figs/distances_early.eps}
\hfill
\includegraphics[height=8cm]{figs/angles_early.eps}
\caption
{
Distances (left) and angles (right) between the ASK classes of the early-type branch.
}
\label{figEarly}
\end{figure}

\begin{figure}
\includegraphics[height=.9\textheight]{figs/distances_late.eps}
\hfill
\includegraphics[height=.9\textheight]{figs/angles_late.eps}
\caption
{
Distances (left) and angles (right) between the ASK classes of the late branch.
}
\label{figLate}
\end{figure}

\begin{figure}
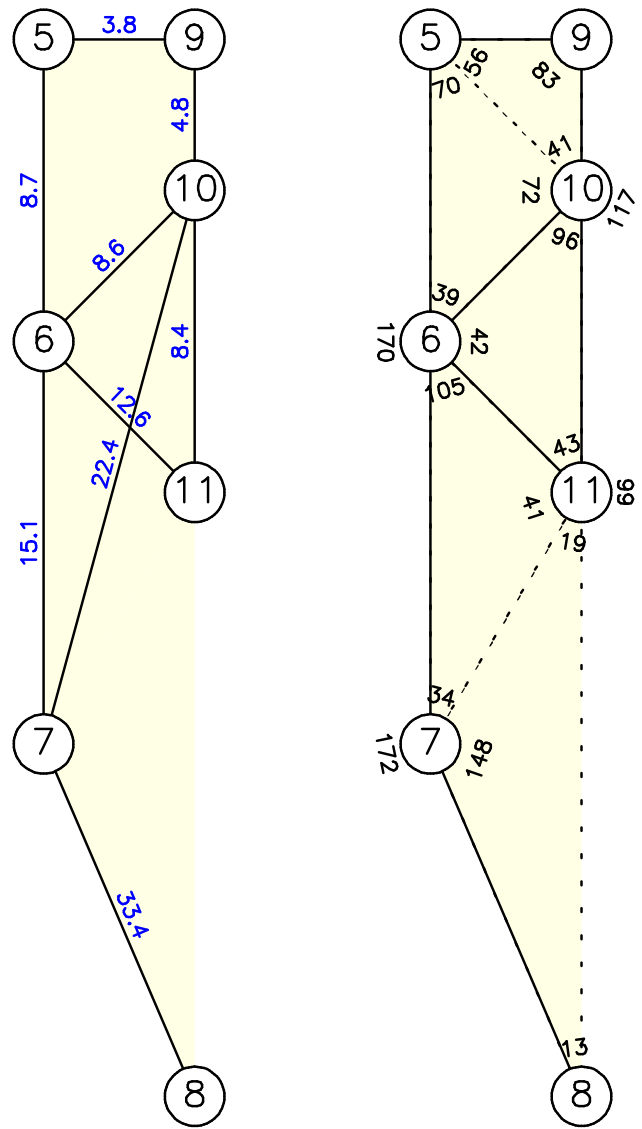

\includegraphics[height=15cm]{figs/distances_active.eps}
\hfill
\includegraphics[height=15cm]{figs/angles_active.eps}
\caption
{
Distances (left) and angles (right) between the ASK classes of the active branch.
}
\label{figActive}
\end{figure}

Given the large amount of data plotted in Figure~\ref{figSequence}, here we represent the early, late, and active branches of our spectroscopic sequence in Figures~\ref{figEarly},~\ref{figLate}, and~\ref{figActive}, respectively.
For the sake of clarity, distances and angles between the different ASK classes are shown in separate panels.

\end{document}